\documentclass[conference]{IEEEtran}
\IEEEoverridecommandlockouts
\usepackage{cite}
\usepackage{amsmath,amssymb,amsfonts}
\usepackage{algorithmic}
\usepackage{graphicx}
\usepackage{textcomp}
\usepackage{xcolor}
\def\BibTeX{{\rm B\kern-.05em{\sc i\kern-.025em b}\kern-.08em
    T\kern-.1667em\lower.7ex\hbox{E}\kern-.125emX}}
\begin{document}

\title{Hybrid Event-triggered Control of Nonlinear System with Full State Constraints and Disturbance\\
}

\author{\IEEEauthorblockN{1\textsuperscript{st} Ziming Wang}
\IEEEauthorblockA{\textit{Systems Hub, Robotics and Autonomous Systems Thrust} \\
\textit{The Hong Kong University of Science and Technology (Guangzhou)}\\
Guangzhou 511458, China \\
	ORCID: 0000-0001-7000-9578}
\and

}

\maketitle

\begin{abstract}
This article focuses on the problem of adaptive tracking control for a specific type of nonlinear system that is subject to full-state constraints via a hybrid event-triggered control (HETC) strategy. With the auxiliary system, we proposed a 'log' function to deal with the full-state constraint. Additionally, a disturbance observer (DO) is constructed to handle the unmeasurable external disturbance. Then, by employing radial basis function neural networks (RBFNNs) and a first-order differentiator, an opportune backstepping design procedure is given to avoid the problem of "explosion of complexity". The HETC strategy, including the fixed and relative threshold, is presented to provide more flexibility in balancing the system performances and network burdens. Finally, to demonstrate the effectiveness of the aforementioned control scheme, a simulation example is presented to validate its effectiveness.
\end{abstract}


\section{Introduction}
For decades, Adaptive control of nonlinear systems has become a research hotspot in the field due to its high intelligence, strong robustness, and wide application range. A plethora of controller design approaches have been extensively proposed and discussed \cite{intro01}\cite{intro02}\cite{intro04}. Among these methods, the adaptive backstepping approach is a substantial and effective innovation in nonlinear control. In practical industrial control settings, it is undeniable that certain uncontrollable factors can have detrimental effects on overall control performance. These factors encompass uncertainties in system dynamics, limitations in load calculation capacity, potential resource scarcity, and both external and internal disturbances. Consequently, researchers are confronted with the following challenges: How can an adaptive controller be designed to effectively address resource scarcity, detect and compensate for unknown disturbances, and accommodate the limitations imposed by system complexity? The aim of this article is to develop a control approach that can be applied to a wider range of control scenarios.\par
Over the past few decades, the rigorous handling of constraints in nonlinear systems has emerged as a fashionable research topic. Constraints are pervasive in physical systems and can be observed as physical stoppages, saturation, as well as performance and safety specifications, among other manifestations. The violation of these constraints during the operation of such systems can result in the degradation of performance or damage to the system. Considering the challenging control problem of system constraints, \cite{intro1} presented a nonlinear mapping-based control design to address the strict-feedback nonlinear systems with output constraints. This method was extended to stochastic pure-feedback nonlinear systems in \cite{intro2}. Furthermore, \cite{muben} combined the one-to-one nonlinear mapping with the adaptive neural dynamic surface control. However, the aforementioned design methods were inadequate in addressing control problems associated with state constraints and dynamic disturbance.\par
Generally speaking, in the realm of control systems, the complexity and potential adverse effects of mismatching disturbances make it crucial to pursue improved disturbance performance for achieving high-accuracy control. However, directly measuring these disturbances is often impractical or costly. To overcome this challenge, one approach is to estimate the disturbance (or its influence) using measurable variables. By utilizing this estimated disturbance, a control action can be implemented to compensate for its effects. This intuitive solution allows for the effective handling of unmeasurable disturbances by utilizing available information to estimate them. A promising method in this regard is disturbance observer-based control \cite{Do1}\cite{Do2}\cite{Do3}\cite{add2}. This method leverages available information to estimate unmeasurable disturbances, enabling effective compensation for their influence. In this article, the observer of external disturbance will be used in state-constrained nonlinear systems.\par
At the early stage, a time-triggered control scheme is mainly used to solve the communication problems. In order to effectively address the issue of unwarranted depletion of system resources, an alternative control scheme, specifically known as event-triggered control (ETC), is designed. \cite{chufa} proposed design methodologies that utilize both the fixed threshold strategy and the relative threshold strategy. Notably, their approach focuses on designing both the adaptive controller and the triggering event simultaneously. This simultaneous design eliminates the need for the input-to-state stability assumption, making the control system more robust and efficient. In contrast to conventional time-triggered controllers, event-triggered controllers offer significant advantages in terms of reducing network resource utilization. Nevertheless, the aforementioned strategies possess distinct advantages in terms of enhancing trace performance and optimizing resource utilization, respectively. The designed HETC strategy, combining the strengths of both of these strategies, is presented to provide greater flexibility in achieving a balance between system performance and network burdens.\par
Motivated by the aforementioned discussions and drawing upon the controller design concepts presented in \cite{muben}, this article introduces an adaptive hybrid event-triggered control strategy for a nonlinear system with full state constraints and unknown external disturbance. The main contributions of this work can be emphasized as follows.\par
1) As compared to the previous results in \cite{con1} and \cite{con2} on the full-state constraint, an auxiliary system is considered to transform a system with constraints into a novel one. Hence, the new challenge for the design of the disturbance observer necessitates further refinement, and the ability of the proposed controller to handle interference is significantly enhanced.\par
2) With the help of the first-order differentiator, the tedious analytical computations and the issue of “explosion of complexity” encountered in the conventional backstepping method are effectively avoided.\par
3) In contrast to the traditional event-triggered control scheme discussed in \cite{con3}\cite{con4}\cite{add1}, the proposed hybrid event-triggered control strategy with the switching boundary can not only obtain a reasonable update interval but also avoid the excessive large impulse.\par
The remainder of this paper is structured as follows. Sec. II continues with a short introduction to the problem formulation and basic assumptions. Sec. III proposes the adaptive tracking controller and discusses the stability. Next, simulation results are performed in Sec. IV. Finally, the conclusion is drawn in Sec. V.

\section{Problem Formulation}
We consider a specific class of nonlinear, time-varying systems with two designed parameters $\Delta_1$ and $\Delta_2$:
\begin{equation}
	\label{deqn_ex1a}
	\begin{cases}
		\dot{\zeta} = p(\zeta,w,t) \\
		\dot{w}_i = f_i(\overline{w}_i,w_{i+1})+d_i(\zeta,w,t)\quad 1\leq i\leq n-1  \\
		\dot{w}_n = f_n(\overline{w}_n,u)+d_n(\zeta,w,t)\quad n\geq2  \\
		y=w{_1}
	\end{cases}
\end{equation}
where $\zeta\in R$ denotes the dynamic; $w_n$, $u\in R$ and $y\in R$ are the state vector, the input and the output, respectively, $w_i$ satisfies: $-\Delta_{i1}<w{_i}<\Delta_{i2}$; $f_i$ and $d_i$, $i=1,...,n$ are unknown nonlinear functions, $d_i$ represents the uncertain disturbance.

\subsection{Full State Constraints}
In the nominal control design, similar to \cite{muben}, we introduce the formulated full state constraints as:
\begin{equation}
	\begin{cases}
		\varpi{_i}=log\frac{\Delta_{i1}+w{_i}}{\Delta_{i2}-w{_i}}\\
		\dot{\varpi{_i}}=\Omega{_i}(\varpi{_i})\dot{w{_i}}
	\end{cases}
\end{equation}
where $\Omega{_i}(\varpi{_i})=\frac{e{^{w{_i}}}+e{^{-w{_i}}}+2}{\Delta_{i1}+\Delta_{i2}}$, $i=1,...,n$.

Applying the following auxiliary system:
\begin{equation}
	\label{deqn_ex1a}
	\begin{cases}
		F_i(\overline{\varpi}_{i+1})=\Omega_i(\varpi_i)f_i(\overline{w}_i,w_{i+1})-\varpi_{i+1}  \\
		F_n(\overline{\varpi}_n)=\Omega_n(\varpi_n)f_n(\overline{w}_n,0)  \\
	\end{cases}
\end{equation}
\begin{equation}
	\label{deqn_ex1a}
	D_i(\zeta,\overline{\varpi}_n,t)=d_i(\zeta,w,t)\quad i=1,...,n
\end{equation}

We convert the system (1) into the following constrained system:
\begin{equation}
	\label{deqn_ex1a}
	\begin{cases}
		\dot{\zeta}=p(\zeta,w,t) \\
		\dot{\varpi}_1=F_1(\varpi_1,\varpi_2)+\Omega_1(\varpi_1)D_1(\zeta,\overline{\varpi}_n,t)+\varpi_2\\
		...\\
		\dot{\varpi}_{n-1}=F_{n-1}(\overline{\varpi}_{n})+\Omega_{n-1}(\varpi_{n-1})D_{n-1}(\zeta,\overline{\varpi}_n,t)+\varpi_n\\
		\dot{\varpi}_n=F_n(\overline{\varpi}_n)+\Omega_n(\varpi_n)u+\Omega_n(\varpi_n)D_n(\zeta,\overline{\varpi}_n,t)\\
	\end{cases}
\end{equation}
where $\overline{\varpi}_i=[\varpi_1,\varpi_2,...,\varpi_i]^T,i=1,2,...,n$

\subsection{Disturbance Observer}
Based on the RBFNNs, the unknown function $F_i(\overline{\varpi}_{i+1})$ over a compact set $F{_x}\subset R{^q}$ can be estimated as:
\begin{equation}
	\label{deqn_ex1a}
	F_i(\overline{\varpi}_{i+1})={W^*_i}^TP_i(\overline{\varpi}_{i+1})+\varGamma(\overline{\varpi}_{i+1})
\end{equation}
where $P_i(\overline{\varpi}_{i+1})=[P{_{i1}}(\overline{\varpi}_{i+1}),...,P{_{im}}(\overline{\varpi}_{i+1})]^T$ is the known function vector with $m>1$ being the RBFNNs node number; $\varGamma(\overline{\varpi}_{i+1})$ is the approximation error. In this brief, $W^*_i$ is chose as $W^*=arg\mathop{\min}\limits_{W\in R{^N}}[\mathop{\sup}\limits_{\overline{\varpi}_{i+1} \in F{_x}}|F_i(\overline{\varpi}_{i+1})-W{^T}P_i(\overline{\varpi}_{i+1})|]$.

To handle the dynamic disturbance, the following DO is suggested with an intermediate variable $\mu_i$ as:
\begin{equation}
	\label{deqn_ex1a}
	\hat{\mu}_i=\hat{D}_i-m_i\varpi_i\quad i=1,...,n
\end{equation}
where $m_i \in R$ is a positive designed constant.

Then, it gives:
\begin{equation}
	\begin{cases}
		\dot{\widetilde{\mu}}{_i}=\dot{D}_i(\zeta,\overline{\varpi}_n,t)-m_i[{\widetilde{W}_i}^TP_i(\overline{\varpi}_{i+1})+\Omega_i(\varpi_i)\widetilde{\mu}_i] \\
		\dot{\widetilde{\mu}}{_n}=\dot{D}_n(\zeta,\overline{\varpi}_n,t)-m_n[{\widetilde{W}_n}^TP_n(\overline{\varpi}_n)+\Omega_n(\varpi_n)\widetilde{\mu}_n]
	\end{cases}
\end{equation}
where $\widetilde{\mu}_i=\mu_i-\hat{\mu}_i$.

\subsection{Hybrid event-triggered control}
The HETC is constructed as
\begin{equation}
	\label{deqn_ex1a}
	k(t)=v(t)-u(t)
\end{equation}
\begin{equation}
	\label{deqn_ex1a}
	u(t)=v(t_s)\quad \forall t\in[t_s, t_{s+1})
\end{equation}
\begin{equation}
	\label{equa:001}
	t_{s+1}=\left\{
	\begin{aligned}
		inf\{t\in R||k(t)|\geq\varUpsilon|u(t)|+\varPhi\} &, \mbox{if }|u(t)|\geq T \\
		inf\{t\in R||k(t)|\geq \varPsi\} \quad\quad\quad\quad&, \mbox{if }|u(t)|<T \\
	\end{aligned}
	\right.\quad
\end{equation}
where $t_s, t_{s+1}\in Z^+$, $0<\varUpsilon<1$, $\Theta$, $\varPhi$ and $\varPsi$ are all designed positive constants. $k(t)$ stands for the error of measurement, while $T$ denotes the switching boundary.

The control protocol is formulated as
\begin{equation}
	\label{deqn_ex1a}
	v(t)=-(1+\varUpsilon)(\alpha_ntanh(\frac{z_n\alpha_n}{H})+Itanh(\frac{z_nI}{H}))
\end{equation}
where $H>0$ and $I>\varPhi/(1-\varUpsilon)$ are all designed constants, and $z_n$, the tracking errors, will be introduced later.

$Remark\quad1:$ 
In this section, we propose a hybrid threshold control strategy, which offers a viable approach to optimizing system performance while effectively managing communication constraints. Concretely, different from \cite{chufa}, the control signal is considered to satisfy the condition $|u(t)|\geq T$. In such cases, the system implements the relative threshold strategy in order to achieve precise control. the controller is afforded enhanced protection, thereby mitigating the risk of damage caused by the abrupt occurrence of a large signal shock. This approach effectively minimizes the potential adverse effects of such shocks on the controller, ensuring its sustained functionality and integrity. However, when the value of the control signal $|u(t)|<T$, the system transitions to the fixed threshold strategy. This strategy facilitates the maintenance of measurement error within a bounded range, thereby ensuring not only specific system performance but also the optimal utilization of resources.

$Assumption\quad1:$ There exist some restrictions with positive constants $\varOmega$ and $\Theta$: $|\Omega_i(\varpi_i)|\leq \varOmega$, $|\dot{D}_i|^2+|D_i|^2\leq|\Theta|^2$ and $f_n(\overline{w}_n,u)=f_n(\overline{w}_n,0)+u$, $i=1,...,n$.

$Assumption\quad2 ${\emph{\cite{muben}\cite{ass2}:} The unmodeled dynamics $\zeta$ is deemed to exhibit exponential input-state-practical stability (exp-ISpS). And there exist unknown nonnegative continuous function $\Xi_{i1}(\cdot)$ and nondecreasing continuous function $\Xi_{i2}(\cdot)$ such that:
	\begin{equation}
		\begin{aligned}
			|d_i(\zeta,w,t)|\leq\Xi_{i1}(||\overline{w}_i||)+\Xi_{i2}(||\zeta||) \\
			\quad\quad\quad\forall (\zeta,w,t)\in R^{n0}\times R^n\times R_+
		\end{aligned}
	\end{equation}
	where $\Xi_{i2}(0)=0, i=1,...,n$.
	
	\emph{Lemma \textsl{1\cite{muben} :}}
	If V is an exp-ISpS Lyapunov function for a system $\dot{\zeta}=p(\zeta,w,t)$, then, for any initial instant $t_0 >0$, any constant $\overline{\wp}\in (0,\wp)$, any initial condition $\zeta_0=\zeta(t_0)$, $\aleph_0>0$, for any continuous function $\overline{\aleph}$ such that $\overline{\aleph}(|w_1|)\geq \aleph(|w_1|)$, there exists a finite $T_0=max\{0,log[(V(\zeta_0)/\aleph_0)]/(\wp-\overline{\wp})\}\geq0$, a nonnegative function $K(t_0,t)$, a constant $\overline{d}\geq0$, and a signal with all $t\geq t_0$ is described by
	\begin{equation}
		\dot{\aleph}=-\overline{\wp}\aleph+\overline{\aleph}(|w_1|)+\overline{d}, \aleph(t_0)=\aleph_0
	\end{equation}
	such that $K(t_0,t)=0$ for $t\geq t_0+T_0$, and the inequality $V(\zeta)\leq \aleph(t)+K(t_0,t)$ holds with $K(t_0,t)=max\{0,e^{-\wp(t-t_0)}V(\zeta_0)-e^{-\overline{\wp}(t-t_0)}\aleph_0\}$, where $log(\bullet)$ represents the natural logarithm of $\bullet$.
	
	\emph{Lemma \textsl{2\cite{con2} :}}
	For any constants $\vartheta_1,\vartheta_2\in R$, the following holds
	\begin{equation}
		\label{deqn_ex1a}
		0\leq |\vartheta_1|-\vartheta_1tanh(\frac{\vartheta_1}{\vartheta_2})\leq0.2785\vartheta_2
	\end{equation}

	\section{Adaptive Control Design}
	Define a notation $\overline{z}_i$, an auxiliary variable $\varphi_i$ and an unknown function $c_i(C_i)$:
	\begin{equation}
		\overline{z}_i=[z_1,z_2,...,z_i]^T
	\end{equation}
	\begin{equation}
		\varphi_i=max\{||W^*_i||^2\}, i=1,...,N
	\end{equation}
	\begin{equation}
		\begin{aligned}
			\label{deqn_ex1a}
			c_i(C_i)=P_i(\overline{\varpi}_{i})+\Omega^2_i(\varpi_i)z_i[\Xi_{i1}(|w_i|)\\+\Xi_{i2}(\overline{\alpha}^{-1}(\aleph+d_0))]^2
		\end{aligned}
	\end{equation}
	where $d_0$ is a positive constant, $C_i=[\overline{\varpi}_{i+1}, z_i, \aleph]\in R^{i+3}$, $W^*_i$ is expressed as an ideal constant weight vector in RBFNNs and define $\widetilde{\varphi}_i=\varphi_i-\hat{\varphi}_i$,  $\hat{\varphi}_i$ is the estimation of $\varphi_i$, $i=1,...,n$. Using RBFNNs to approximate $c_i(C_i)$ as $c_i(C_i)={W^*_i}^TP_i(C_i)+\varGamma(C_i)$, $i=1,...,n-1$.
	
	The preceding description outlines the systematic configuration procedure, whereas the block diagram illustrating the adaptive control framework with HETC is depicted in Fig.1. The subsequent section delves into the detailed theoretical derivation.
	\begin{figure}[!t]
		\centering
		\includegraphics[height=4.5cm,width=9cm]{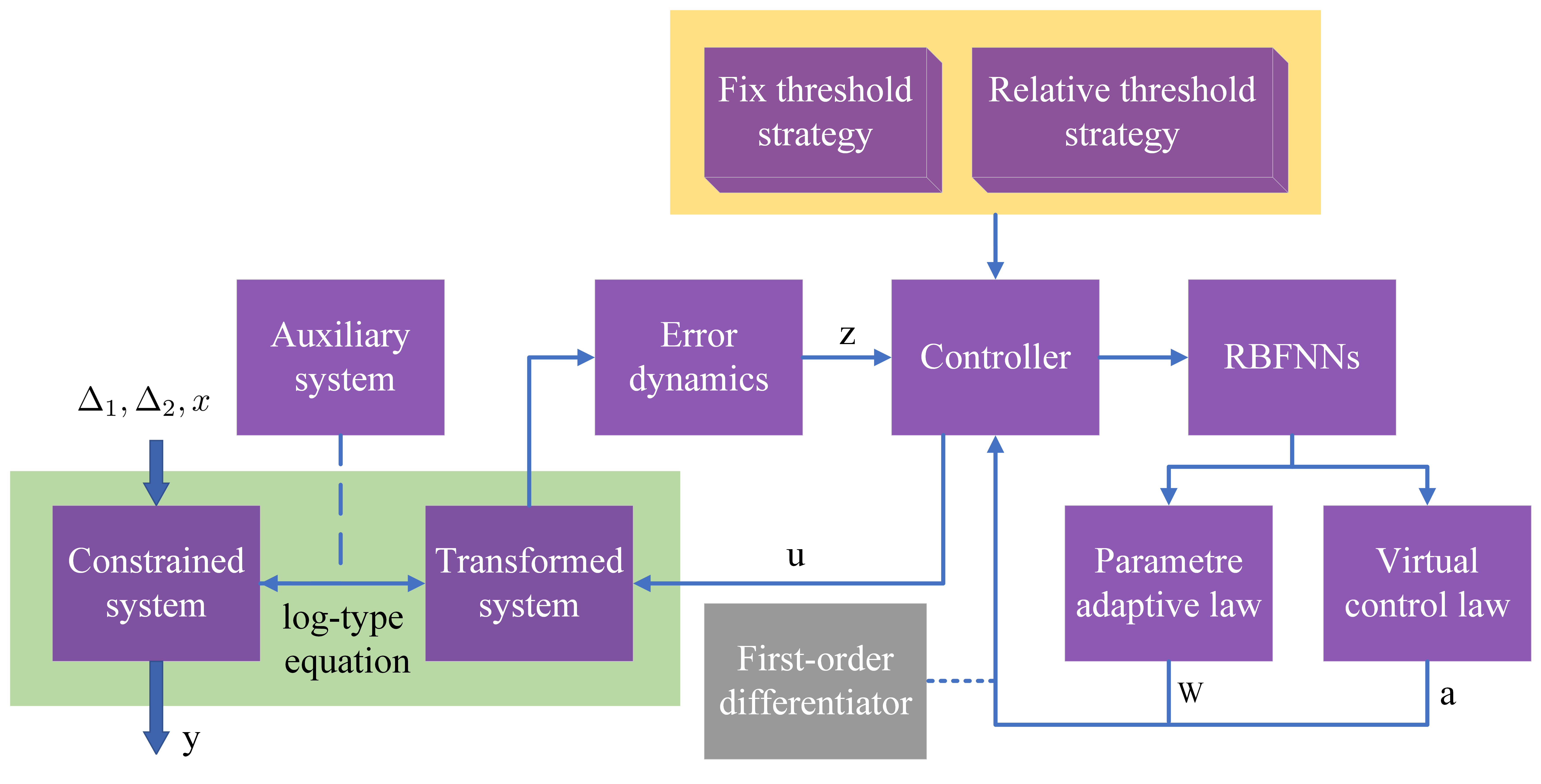}\\
		\caption{Block diagram of the adaptive control framework with HETC.}
	\end{figure}
	
	$Step\quad1:$
	We define ${y}_r=log\frac{\Delta_1+w_r}{\Delta_2-w_r}$, where $\varpi_r$ is the  desired reference signal. Then, we have tracking error $z{_1}$:
	\begin{equation}
		z{_1}=\varpi_1-{y}_r
	\end{equation}
	
	Construct the virtual control $\alpha_1$ and the adaptive law $\dot{\hat{W}}_1$ with designed constants $\xi_1, \lambda_1, a_1, e_1,>0$ as follows:
	\begin{equation}
		\begin{aligned}
			\label{deqn_ex1a}
			\alpha_1=-\xi_1z_1-\frac{z_1}{2a^2_1}\hat{\varphi}||P_1(C_1)||^2-\Omega_1(\varpi_1)\hat{D}_1(\zeta,\overline{\varpi}_n,t)+\dot{\hat{y}}_r
		\end{aligned}
	\end{equation}
	\begin{equation}
		\begin{aligned}
			\label{deqn_ex1a}
			\dot{\hat{W}}_1=-\lambda_1z_1m_1P_1(\overline{\varpi}_2)-e_1\hat{W}_1	
		\end{aligned}
	\end{equation}
	
	Select the Lyapunov function:
	\begin{equation}
		\label{deqn_ex1a}
		V_1=\frac{\widetilde{\mu}^2_1}{2}+\frac{z^2_1}{2}+\frac{1}{2\psi}\widetilde{\varphi}^2+\frac{1}{2\lambda_1}\widetilde{W}^T_1\widetilde{W}_1
	\end{equation}
	
	From Young's inequality, we have that
	\begin{equation}
		z_1\varGamma(C_1)\leq\frac{z^2_1}{2}+\frac{(\varGamma(C_1))^2}{2}     
	\end{equation}
	\begin{equation}
		\widetilde{\mu}_1\dot{D}_1(\zeta,\overline{\varpi}_n,t)\leq\frac{\widetilde{\mu}^2_1}{2}+\frac{\dot{D}_1(\zeta,\overline{\varpi}_n,t)^2}{2}    
	\end{equation}
	\begin{equation}
		z_1m_1P_1(\overline{\varpi}_2)\leq\frac{z^2_1}{2}+\frac{m^2_1Q^2_1}{2}
	\end{equation}
	\begin{equation}
		-m_1\widetilde{W}^T_1P_1(\overline{\varpi}_2)\leq\frac{1}{2}\widetilde{W}^T_1W_1+\frac{m^2_1Q^2_1}{2}   
	\end{equation}
	\begin{equation}
		-m_1\Omega_1(\varpi_1)\widetilde{\mu} ^2_1\leq\frac{1}{4}\widetilde{\mu} ^2_1+m_1^2\varOmega^2\widetilde{\mu} ^2_1
	\end{equation}
	where $Q_i=||P_i(\varpi_{i+1})||$, $i=1,...,n-1$.
	
	Therefore, based on (23) to (27), the time derivative of $V_1$ is expressed as follows:
	\begin{equation}
		\begin{aligned}
			\label{deqn_ex1a}
			\dot{V}_1&=z_1\dot{z}_1+\widetilde{\mu}_1\dot{\widetilde{\mu}}_1-\frac{1}{\psi}\widetilde{\varphi}\dot{\hat{\varphi}}-\frac{1}{\lambda_1}\widetilde{W}^T_1\dot{\hat{W}}_1\\
			&\leq z_1z_2+(1-\xi_1)z^2_1+[\frac{3}{4}+m_1^2\varOmega^2]\widetilde{\mu}^2_1+\frac{a^2_1}{2}\\
			&\quad\quad+\frac{|\varGamma(C_1)|^2}{2}+\frac{\Theta^2}{2}-\frac{1}{\psi}\widetilde{\varphi}\dot{\hat{\varphi}}+\frac{e_1}{\lambda_1}\widetilde{W}^T_1\hat{W}_1+\frac{1}{4}\\
			&\quad\quad+\frac{z_1}{2a^2_1}\widetilde{\varphi}||P_1(C_1)||^2+m^2_1Q^2_1+\frac{1}{2}\widetilde{W}^T_1W_1	
		\end{aligned}
	\end{equation}
	
	$Step\quad i\quad (2\leq i\leq n-1):$
	Select the $i$-th Lyapunov function:
	\begin{equation}
		\label{deqn_ex1a}
		V_i=\frac{\widetilde{\mu}^2_i}{2}+\frac{z^2_i}{2}+\frac{1}{2\psi}\widetilde{\varphi}^2+\frac{1}{2\lambda_i}\widetilde{W}^T_i\widetilde{W}_i
	\end{equation}
	where $z_i=m_i-\alpha_{i-1}$.
	
	In order to bypass the tedious analytical calculation of the virtual law $\dot{\alpha}_{i-1}$, the adoption of the subsequent first-order differentiator, as proposed by \cite{muben}, is employed to estimate $\dot{\alpha}_{i-1}$
	\begin{equation}
		\begin{cases}
			\label{deqn_ex1a}
			\dot{\delta}_{i0}=\sigma_{i}=\delta_{i1}-\epsilon_{i0}|\delta_{i0}-\alpha_{i-1}|^{\frac{1}{2}}sign(\delta_{i0}-\alpha_{i-1}) \\
			\dot{\delta}_{i1}=-\epsilon_{i1}sign(\delta_{i1}-\sigma_{i})
		\end{cases}
	\end{equation}
	where $\epsilon_{i0}$ and $\epsilon_{i1}$ are designed positive constants. $\delta_{i0},\delta_{i1}$, and $\sigma_{i}$ are the states of the system (30), $i=2,...,n$.
	
	According to (30) and \cite{erjie}, one gets:
	\begin{equation}
		\dot{\alpha}_{i-1}=\sigma_{i}+\varLambda_{i-1}  \quad i=2,...,n  
	\end{equation}
	where $\varLambda_{i-1}$ represents the estimation error of the first-order sliding-mode differentiator. According to \cite{erjie}, we know that $|\varLambda_{i-1}|\leq\overline {\varLambda}_{i-1}$ with $\overline {\varLambda}_{i-1}>0$.
	
	$Remark\quad2:$ 
	The conventional backstepping method is known to have a problem of "explosion of complexity." This is due to the incorporation of the RBFNNs in our backstepping control scheme allows us to effectively handle system uncertainty. However, this inclusion introduces the need to calculate derivatives of the radial basis functions, which in turn increases the computational burden at each step of the design. This issue is addressed in this paper by introducing a first-order sliding-mode differentiator. By utilizing this differentiator, the derivatives of virtual control laws can be accurately computed, thus alleviating the computational load associated with backstepping control design.
	
	Select the virtual control $\alpha_i$ and the adaptive law $\dot{\hat{W}}_i$ as follows:
	\begin{equation}
		\begin{aligned}
			\label{deqn_ex1a}
			\alpha_i=-\xi_iz_i-\frac{z_i}{2a^2_i}\hat{\varphi}||P_i(C_i)||^2-\Omega_i(\varpi_i)\hat{D}_i(\zeta,\overline{\varpi}_n,t)+\sigma_i
		\end{aligned}
	\end{equation}
	\begin{equation}
		\begin{aligned}
			\label{deqn_ex1a}
			\dot{\hat{W}}_i=-\lambda_iz_im_iP_i(\overline{\varpi}_{i+1})-e_i\hat{W}_i	
		\end{aligned}
	\end{equation}
	where $\xi_i, \lambda_i, a_i, e_i>0$ are designed constants.
	
	From Young's inequalities, similar to (23)-(27) and $-z_i\varLambda_{i-1}\leq\frac{z^2_i}{2}+\frac{(\overline{\varLambda}_{i-1})^2}{2}$, $i=2,...,n$, we have that
	\begin{equation}
		\begin{aligned}
			\label{deqn_ex1a}
			\dot{V}_{i}&\leq z_iz_{i+1}+(\frac{3}{2}-\xi_i)z^2_i+[\frac{3}{4}+m^2_i\varOmega^2]\widetilde{\mu}^2_i+\frac{a^2_i}{2}+\frac{1}{4}\\
			&\quad\quad+\frac{|\varGamma(C_i)|^2}{2}+\frac{\Theta^2}{2}-\frac{1}{\psi}\widetilde{\varphi}\dot{\hat{\varphi}}+\frac{e_i}{\lambda_i}\widetilde{W}^T_i\hat{W}_i+\frac{(\overline{\varLambda}_{i-1})^2}{2}\\
			&\quad\quad+\frac{z_i}{2a^2_i}\widetilde{\varphi}||P_i(C_i)||^2+m^2_iQ^2_i+\frac{1}{2}\widetilde{W}^T_iW_i
		\end{aligned}
	\end{equation}
	
	$Step\quad n:$
	Select the Lyapunov function as follows:
	\begin{equation}
		\label{deqn_ex1a}
		V_n=\frac{\widetilde{\mu}^2_n}{2}+\frac{z^2_n}{2}+\frac{1}{2\psi}\widetilde{\varphi}^2+\frac{1}{2\lambda_n}\widetilde{W}^T_n\widetilde{W}_n
	\end{equation}
	where $z_n=\varpi_n-\alpha_{n-1}$.
	
	And $\dot{\hat{\varphi}}$ is given as:
	\begin{equation}
		\begin{aligned}
			\label{deqn_ex1a}
			\dot{\hat{\varphi}}=\frac{z^2_n}{2a^2_0}||P_n(z_n)||^2-\tau\hat{\varphi}
		\end{aligned}
	\end{equation}
	where $\tau$ and $a_0$ are designed constants.
	
	Select the virtual control $\alpha_n$ and the adaptive law $\dot{\hat{W}}_n$ as follows:
	\begin{equation}
		\begin{aligned}
			\label{deqn_ex1a}
			\alpha_n&=\frac{1}{\Omega_n(\varpi_n)}[-\xi_nz_n-\frac{z_n}{2a^2_n}\hat{\varphi}||P_n(C_n)||^2-\Omega_n(\varpi_n)\\
			&\quad\quad\hat{D}_n(\zeta,\overline{\varpi}_n,t)+\sigma_n]
		\end{aligned}
	\end{equation}
	\begin{equation}
		\begin{aligned}
			\label{deqn_ex1a}
			\dot{\hat{W}}_n=-\lambda_nz_nm_nP_n(\overline{\varpi}_{n})-e_n\hat{W}_n	
		\end{aligned}
	\end{equation}
	where $\xi_n, \lambda_n, a_n, e_n>0$ are design constants.
	
	$Case\quad1:$ $|u(t)|\geq T$. Considering the time-varying continuous functions $\kappa_1(t)$ and $\kappa_2(t)$,  $|\kappa_1(t)|\leq1$, $|\kappa_2(t)|\leq1$, $\forall t\in[t_s,t_{s+1})$, and referring to equation (11), one can express $v(t)$ as
	$v(t)=(1+\kappa_1(t)\varUpsilon)u(t)+\kappa_2(t)\varPhi$. Then one can derive the expression for $u(t)$ as $u(t)=(v(t)/(1+\kappa_1(t)\varUpsilon))-(\kappa_2(t)\varPhi/(1+\kappa_1(t)\varUpsilon))$.
	
	one has:
	\begin{equation}
		\begin{aligned}
			\label{deqn_ex1a}
			\frac{z_nv(t)}{1+\kappa_1(t)\varUpsilon}\leq \frac{z_nv(t)}{1+\varUpsilon}\\
			|\frac{\kappa_2(t)\varPhi_1}{1+\kappa_1(t)}|\leq \frac{\varPhi}{1-\varUpsilon}
		\end{aligned}
	\end{equation} 
	
	Hence, similar to $ist$ and adopting the relative threshold strategy, it holds that
	\begin{equation}
		\begin{aligned}
			\label{deqn_ex1a}
			\dot{V}_{n}&\leq (\frac{3}{2}-\xi_n)z^2_n+[\frac{3}{4}+m^2_n\varOmega^2]\widetilde{\mu}^2_n+\frac{a^2_n}{2}+\frac{1}{4}+\frac{\Theta^2}{2}\\
			&\quad\quad+\frac{|\varGamma(C_n)|^2}{2}-\frac{\tau}{\psi}\widetilde{\varphi}{\hat{\varphi}}+\frac{e_n}{\lambda_n}\widetilde{W}^T_n\hat{W}_n+\frac{(\overline{\varLambda}_{n-1})^2}{2}\\
			&\quad\quad-|z_nI|+|\frac{z_n\varPhi}{1-\varUpsilon}|+0.557\varOmega H+m^2_nQ^2_n+\frac{1}{2}\widetilde{W}^T_nW_n\\
		\end{aligned}
	\end{equation}
	
	$Case\quad2:$ $|u(t)|<T$. By employing a similar approach to the relative threshold strategy in Case 1, combining the same terms, one can easily obtain
	\begin{equation}
		\begin{aligned}
			\label{deqn_ex1a}
			\dot{V}_{n}&\leq (\frac{3}{2}-\xi_n)z^2_n+[\frac{3}{4}+m^2_n\varOmega^2]\widetilde{\mu}^2_n+\frac{a^2_n}{2}+\frac{1}{4}+\frac{\Theta^2}{2}\\
			&\quad\quad+\frac{|\varGamma(C_n)|^2}{2}-\frac{\tau}{\psi}\widetilde{\varphi}{\hat{\varphi}}+\frac{e_n}{\lambda_n}\widetilde{W}^T_n\hat{W}_n+\frac{(\overline{\varLambda}_{n-1})^2}{2}\\
			&\quad\quad-|z_nI|+|\frac{z_n\varPhi}{1-\varUpsilon}|+1.114\varOmega H+m^2_nQ^2_n+\frac{1}{2}\widetilde{W}^T_nW_n\\
		\end{aligned}
	\end{equation}
	
	\itshape Theorem 1:
	\upshape Consider the closed-loop system consisting of the system (1), the controller (37), the adaptation law (38). There exist constants $a_i>0, \epsilon_{i0}>0, \epsilon_{i1}>0$ such that all signals within the closed-loop system are semiglobally uniformly ultimately bounded (SGUUB), and the full state constraints remain unviolated. Meanwhile, the Zeno behaviour is effectively excluded. In addition, $\xi_i, m_i, e_i, \lambda_i, \tau$ satisfy
	\begin{equation}
		\begin{cases}
			\label{deqn_ex1a}
			\xi_i\geq\frac{5}{2}+\frac{\beta}{2}\quad i=1,...,n.\\
			m^2_i\leq-\frac{3}{4\varOmega^2}-\frac{\beta}{2\varOmega^2}\quad i=1,...,n.\\
			\frac{e_i}{\lambda_i}\leq-\frac{1}{2}-\frac{\beta}{2}\quad i=1,...,n.\\
			\tau\leq-\frac{\beta}{2}.\\
			\beta=min\{(\frac{5}{2}-\xi_i),(\frac{3}{4}+m^2_i\varOmega^2),(\frac{e_i}{\lambda_i}+\frac{1}{2}),\frac{n\tau}{\psi}\}
		\end{cases}
	\end{equation}\par
	\itshape Proof:
	\upshape To address the convergence of all signals in the closed-loop system, we establish the Lyapunov function candidate for the entire system as follows:
	\begin{equation}
		\begin{aligned}
			V=\sum\limits_{i=1}^{n} V_i
		\end{aligned}
	\end{equation}
	
	Differentiating $V$ and considering (28), (34), and (40), one obtains:
	\begin{equation}
		\begin{aligned}
			\label{deqn_ex1a}
			\dot{V}&\leq 
			\sum\limits_{i=1}^{n}(\frac{5}{2}-\xi_i)z^2_i+\sum\limits_{i=1}^{n}[\frac{3}{4}+m^2_i\varOmega^2]\widetilde{\Upsilon}^2_i+\sum\limits_{i=1}^{n}(\frac{e_i}{\lambda_i}+\frac{1}{2})\\
			&\quad\quad\widetilde{W}^T_i\hat{W}_i+\frac{n\tau}{\psi}\widetilde{\varphi}\hat{\varphi}+\sum\limits_{i=1}^{n}\frac{a^2_i}{2}+\frac{n}{4}+\frac{n\Theta^2}{2}+\sum\limits_{i=1}^{n}\frac{|\varGamma(C_i)|^2}{2}\\
			&\quad\quad+\sum\limits_{i=2}^{n}\frac{(\overline{\varLambda}_{i-1})^2}{2}+\sum\limits_{i=1}^{n}m^2_iQ^2_i
		\end{aligned}
	\end{equation}
	
	By the fact that:
	\begin{equation}
		\begin{cases}
			\label{deqn_ex1a}
			\widetilde{\varphi}_i\hat{\varphi}_i\leq-\frac{1}{2}\widetilde{\varphi}_i^2+\frac{1}{2}{\varphi}_i^2 \\
			\widetilde{W}^T_i\hat{W}_i\leq-\frac{1}{2}\widetilde{W}^T_i\widetilde{W}_i+\frac{1}{2}W^T_iW_i
		\end{cases}
	\end{equation}
	
	Set:
	\begin{equation}
		\begin{aligned}
			\label{deqn_ex1a}
			\gamma&=\sum\limits_{i=1}^{n}\frac{a^2_i}{2}+\frac{n}{4}+\frac{n\Theta^2}{2}+\sum\limits_{i=1}^{n}\frac{|\varGamma(C_i)|^2}{2}+\sum\limits_{i=2}^{n}\frac{(\overline{\varLambda}_{i-1})^2}{2}\\
			&\quad\quad+\sum\limits_{i=1}^{n}m^2_iQ^2_i+\sum\limits_{i=1}^{n}\frac{1}{2}{\varphi}_i^2+\sum\limits_{i=1}^{n}\frac{1}{2}W^T_iW_i
		\end{aligned}
	\end{equation}
	
	Substituting (42) (45) and (46) into (44), we obtain:
	\begin{equation}
		\begin{aligned}
			\label{deqn_ex1a}
			\dot{V}\leq-\beta V+\gamma
		\end{aligned}
	\end{equation}
	$\dot{V}\leq0$ with the condition of $V=r$ and $\beta>(\gamma/r)$. We can obtain that for any time $t\geq0$ if $V(0)\leq r$ then $V(t)\leq r$. Multiplying (47) by $e^{\beta t}$ yields
	\begin{equation}
		\begin{aligned}
			\label{deqn_ex1a}
			\frac{d}{dt}(V(t)e^{\beta t})\leq e^{\beta t}\gamma
		\end{aligned}
	\end{equation}
	
	Integrating (48) over $[0,t]$, we get
	\begin{equation}
		\begin{aligned}
			\label{deqn_ex1a}
			0\leq V(t)\leq\frac{\gamma}{\beta}+[V(0)-\frac{\gamma}{\beta}]e^{-\beta t}
		\end{aligned}
	\end{equation}
	
	Hence, all signals in the closed-loop system are semiglobally uniformly ultimately bounded. From (49), we obtain $z_i\leq\sqrt{(2\gamma/\beta)+2[V(0)-(\gamma/\beta)]e^{-\beta t}}$. Therefore, $z_1$ as $t\rightarrow\infty$ can be made arbitrarily small.
	
	one has
	\begin{equation}
		\begin{aligned}
			\label{deqn_ex1a}			
			\frac{d}{dt}|k(t)|&=\frac{d}{dt}(k(t)\times k(t))^{\frac{1}{2}}\\&=sign(k(t))\dot{k}(t)\leq|\dot{v}(t)|
		\end{aligned}
	\end{equation}
	
	From (12), we can get
	\begin{equation}
		\begin{aligned}
			\label{deqn_ex1a}			\dot{v}(t)&=-(1+\varUpsilon)(\alpha_ntanh(\frac{z_n\alpha_n}{H})+\alpha_n\frac{\frac{\dot{z}_n\alpha_n}{H}+ \frac{z_n\dot{\alpha}_n}{H}}{cosh^2(\frac{z_n\alpha_n}{H})}+\\&\quad\quad\frac{I^2\dot{z}_n}{Hcosh^2(\frac{z_nI^2}{H})})
		\end{aligned}
	\end{equation}	
	
	Based on the aforementioned discussion, all the closed-loop signals are bounded. It further implies that $|\dot{v}(t)|\leq\varrho$ where $\varrho$ is a positive parameter. According to (10) and (11), $k(t)=0$ and $lim_{t\rightarrow t_{k+1}}k(t)=\varUpsilon|u(t)|+\varPhi$. Then, the lower bound on mutual execution time $t^*$ satisfies $t^*\geq[(\varUpsilon|u(t)|+\varPhi)/\varrho]$ for $\forall t\in [t_k,t_{k+1})$, i.e., the Zeno behavior can be operatively eliminated. A numerical example is given in the next section. 
	
	\section{Validation}
	To demonstrate the effectiveness of the proposed approach, an example can be seen as follows:
	
	Consider the following nonlinear system:
	\begin{equation}
		\begin{aligned}
			\begin{cases}
				\label{deqn_ex1a}
				\dot{\zeta}=-\zeta+x_1^2cos(t)+\frac{1}{5}\\
				\dot{x}_1=x_1+\frac{x_2}{2}+\frac{x_2^3}{3}+d_1\\
				\dot{x}_2=x_1x_2+(\frac{sin(\frac{1}{2}sin(x_1)x_2)u}{5}+\frac{(u^3+\frac{1}{10})}{7}+2d_2)
			\end{cases}
		\end{aligned}
	\end{equation}
	where $d_1=13(\zeta sin(x_1))+1, d_2=\frac{3}{5}cos(\zeta t+x_2-1)\zeta-\frac{1}{10}$, and the errors of two disturbance observer can be defined as $\nu_1=d_1-\mu_1$ and $\nu_2=d_2-\mu_2$. The desired tracking trajectory is $x_r=\frac{3sin(4t)+cos(t)}{10}$.
	
	The design parameters are taken as: $\Delta_{b_{11}}=\Delta_{b_{12}}=2.1, \Delta_{b_{21}}=2, \Delta_{b_{22}}=2.4, a_1=10, a_2=60, \xi_1=150, \xi_2=185, m_1=15, m_2=0.05, e_1=20, e_2=50, \varUpsilon=0.3, \varPhi=1, I=3, H=900, \lambda_1=\lambda_2=1, \epsilon_{10}=2, \epsilon_{20}=2.9, \tau=1.5$. The initial conditions: $x_1(0)=0.1, x_2(0)=-0.1, y_d(0)=0, \varphi(0)=0.5, \delta_0(0)=\delta_1(0)=0.1, \zeta(0)=0, \mu_1(0)=\mu_2(0)=0$. The simulation results are demonstrated in Figs.2-3. From Fig. 2, it is evident that the system achieves satisfactory tracking performance. The intervals of HETC show that the overall triggering times are 977, including 82 triggering times for fixed threshold and 895 triggering times for relative threshold, which is significantly lower than the time sampling strategy of 20000 times. Fig. 2 also indicates that the Zeno phenomenon is successfully avoided. Phase portrait of two disturbance observation errors is shown in Fig. 3, which synchronously shows the curve of the control signal $u(t)$ and the estimated parameter $\varphi$. 
	
	\begin{figure}[!t]
		\centering
		\includegraphics[height=6.5cm,width=9.5cm]{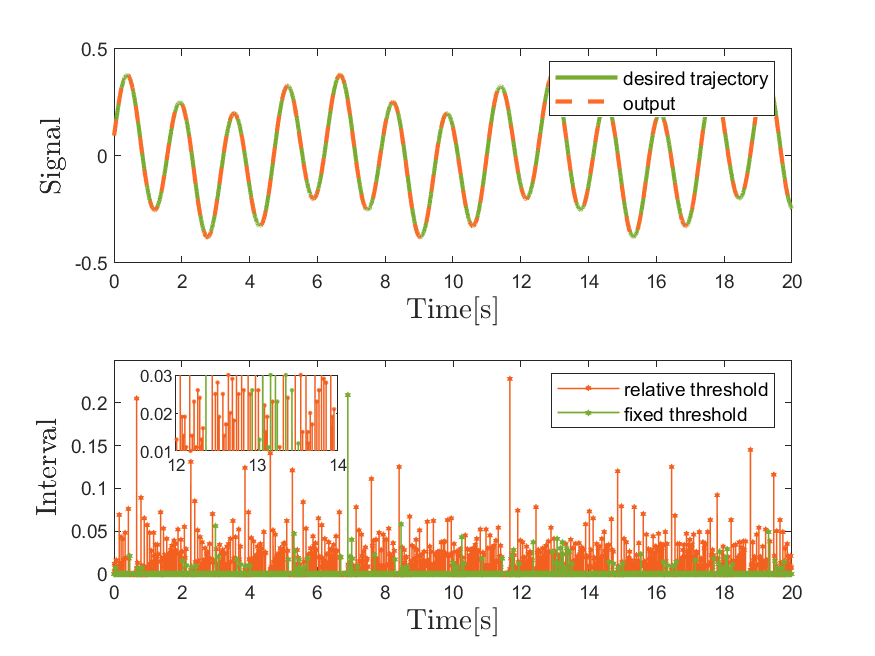}\\
		\caption{Desired trajectories $m_r$ and outputs $m_1$, the interval of HETC.}
	\end{figure}
	
	\begin{figure}[!t]
		\centering
		\includegraphics[height=6.5cm,width=9.5cm]{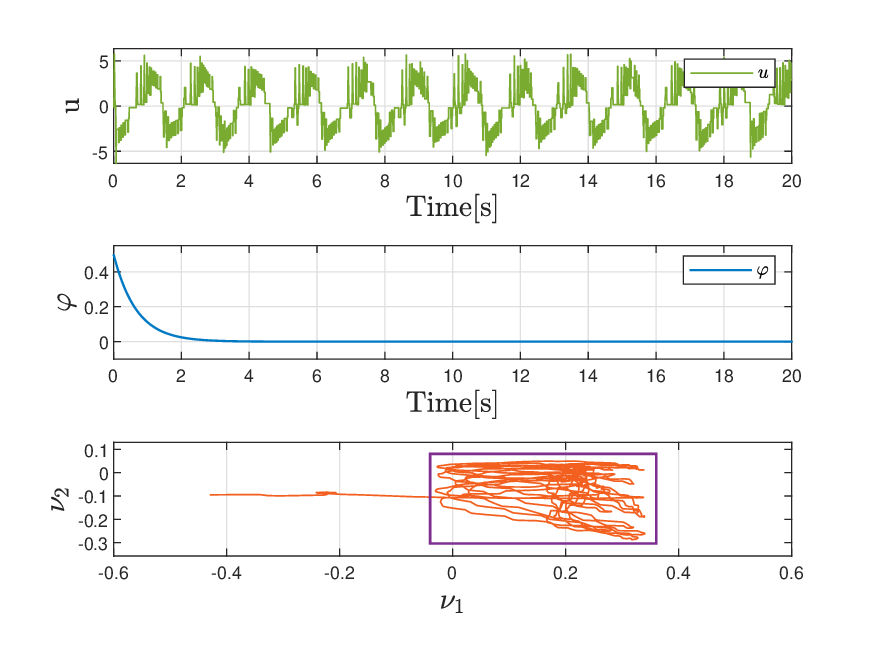}\\
		\caption{Control signal $u(t)$, estimated parameter $\varphi$ and phase portrait of two disturbance observer errors.}
	\end{figure}
	
	\section{Conclusions}
	In this paper, a hybrid event-triggered control strategy is proposed for a full-state-constraint nonlinear system. By introducing a switching boundary, the controller performance is effectively developed and the resource utilization is highly reduced. Moreover, we propose the disturbance observer and the first-order differentiator, and the obtained disturbance estimation information was intelligently utilized to formulate the control laws, thereby enhancing the system's ability to combat external interferences and improve its anti-interference capacity. It is proved that all the closed-loop signals are semiglobally uniformly ultimately bounded. Ultimately, this research has effectively utilized a quintessential example to unveil the inherent availability and unwavering dependability of the proposed adaptive control approach.

\end{document}